\documentclass[apj]{emulateapj}
\usepackage{times}
\usepackage{microtype}

\newcommand{\degg}{\hbox{$^\circ$}}

\newcommand{\xmm}{{\it XMM-Newton}}
\newcommand{\suzaku}{{\it Suzaku}}

\newcommand{\ls}{\mathrel{\hbox{\rlap{\hbox{\lower4pt\hbox{$\sim$}}}\hbox{$<$}}}}
\newcommand{\gs}{\mathrel{\hbox{\rlap{\hbox{\lower4pt\hbox{$\sim$}}}\hbox{$>$}}}}

\begin{document}

\title{Variability of the High Velocity Outflow in the Quasar PDS 456}
\shorttitle{The Outflow in PDS 456}
\shortauthors{Reeves et al.}
\slugcomment{{\sc Accepted to ApJ:} November 1, 2013}

\author{J. N. Reeves\altaffilmark{1,2}, V. Braito\altaffilmark{3}, 
J. Gofford\altaffilmark{1}, S. A. Sim\altaffilmark{4}, E. Behar\altaffilmark{5}, 
M. Costa\altaffilmark{1}, S. Kaspi\altaffilmark{5,6}, G. Matzeu\altaffilmark{1}, 
L. Miller\altaffilmark{7}, P. O'Brien\altaffilmark{8}, T. J. Turner\altaffilmark{2}, 
M. Ward\altaffilmark{9}}

\affil{
$^{1}$Astrophysics Group, School of Physical and Geographical Sciences, Keele 
University, Keele, Staffordshire, ST5 5BG, UK; j.n.reeves@keele.ac.uk\\
$^{2}$Center for Space Science and Technology, University of Maryland Baltimore County, 1000 Hilltop Circle, Baltimore, MD 21250, USA\\
$^{3}$INAF - Osservatorio Astronomico di Brera, Via Bianchi 46 I-23807 Merate (LC), Italy\\
$^{4}$School of Mathematics and Physics, Queen's University Belfast, Belfast BT7 1NN, UK\\
$^{5}$Dept of Physics, Technion, Haifa 32000, Israel\\
$^{6}$School of Physics \& Astronomy and the Wise Observatory, Tel-Aviv University, Tel-Aviv 69978, Israel\\ 
$^{7}$Dept of Physics, University of Oxford, Denys Wilkinson Building, Keble Road, Oxford OX1 3RH, UK\\
$^{8}$Dept of Physics and Astronomy, University of Leicester, University Road, Leicester LE1 7RH, UK\\
$^{9}$Dept of Physics, University of Durham, South Road, Durham DH1 3LE, UK
}

\begin{abstract}

We present a comparison of two \suzaku\ X-ray observations of the nearby (z=0.184), luminous ($L_{\rm bol} \sim 10^{47}$\,erg\,s$^{-1}$) type I quasar, PDS\,456. A new 125\,ks \suzaku\ observation in 2011 caught the quasar during a period of low X-ray flux and with a hard X-ray spectrum, in contrast to a previous 190\,ks 
\suzaku observation in 2007 when the quasar appeared brighter and had a steep ($\Gamma>2$) X-ray spectrum. 
The 2011 X-ray spectrum contains a pronounced trough near 9\,keV in the quasar rest frame, which can be modeled with blue-shifted iron K-shell absorption, most likely from the He and H-like transitions of iron. The absorption trough is observed at a similar rest-frame energy as in the earlier 2007 observation, which appears to confirm the existence of a persistent high velocity wind in PDS 456, at an outflow velocity of $0.25-0.30$\,c. The spectral variability between 2007 and 2011 can be accounted for by variations in a partial covering absorber, increasing in covering fraction from the brighter 2007 observation to the hard and faint 2011 observation. Overall the low flux 2011 observation can be explained if PDS\,456 is observed at relatively low inclination angles through a Compton thick wind, originating from the accretion disk, which significantly attenuates the X-ray flux from the quasar. 
\end{abstract}

\keywords{black hole physics --- quasars: individual: PDS 456 --- X-rays: galaxies}

\section{Introduction}

Outflows are an important phenomenon in AGN and can play a key role in the co-evolution of the massive black hole and the host galaxy. Black holes grow by accretion and strong nuclear outflows can quench this process by effectively shutting off the supply of matter. Thus outflows can provide feedback between the growth of the galactic bulge and the black hole \citep{SR98,King03,DiMatteo05,McMc12}. Recently a number of high column density ($N_{\rm H}\sim 10^{23}$\,cm$^{-2}$), fast ($>0.1c$) outflows have been found in luminous AGN, through observations of blue-shifted absorption from Fe K, at rest-frame energies greater than 7 keV \citep[e.g.][]{Reeves03,R09,Pounds03,Chartas03,Tombesi10,Tombesi11,Gofford13}. These fast outflows may be driven off the accretion disk by either radiation pressure \citep{Proga00,Sim10b} or by magneto-rotational forces \citep{Kato04,Kazanas12,Fukumura10}, or both, a few gravitational radii ($R_{\rm g}$) from the black hole. Under the assumption of quasi-spherical outflow, then the outflow rates derived can be huge, of the order of several solar masses per year \citep{R09,Tombesi11} equivalent to $10^{45} - 10^{46}$\,erg\,s$^{-1}$ in kinetic power. Such outflows are a likely consequence of near-Eddington accretion \citep{KP03}.

PDS 456 is a luminous, low redshift ($z=0.184$) radio-quiet quasar identified in 1997 \citep{Torres97}. The optical and infra-red spectra \citep{Simpson99} show broad Balmer and Paschen lines (e.g. H$\beta$ $\sim3000$\,km~s$^{-1}$ FWHM), strong Fe \textsc{ii}, a hard (de-reddened) optical continuum ($f_{\nu} \propto \nu^{-0.1\pm0.1}$) and one of the strongest `big blue bumps' of any AGN \citep{Simpson99,Reeves00}. It has a de-reddened, absolute magnitude of M$_{B}\approx -27$, making it at least as luminous as the radio-loud quasar 3C\,273, with a bolometric luminosity of $L_{\rm bol}=10^{47}$\,erg\,s$^{-1}$ \citep{Simpson99,Reeves00}, making PDS\,456 one of the most luminous known quasars in the local Universe. 

A 2001 40\,ks XMM-Newton observation of PDS~456 revealed the presence of strong iron K-shell absorption above 7 keV, which could be attributed to a high velocity outflow \citep{Reeves03}, requiring a large column density of highly ionized matter. Furthermore a strong broad absorption line at 1 keV was observed in the RGS  \cite[also see][]{Behar10}. A \suzaku observation in 2007 revealed the presence of two highly significant absorption lines observed at 7.68 and 8.15\,keV \citep[hereafter R09]{R09}, corresponding to 9.08 and 9.66 keV in the quasar rest frame, where  no strong atomic transitions are expected. Furthermore the observed energy of the absorption does not coincide with the expected position of the Fe~\textsc{xxv} or Fe~\textsc{xxvi} resonant ($1s-2p$) absorption lines at $z=0$ (at 6.7 or 6.97 keV) and therefore cannot be associated with local hot gas, such as the WHIM 
\citep{McKernan05}.

Thus a plausible identification of the Fe K absorption lines is with the strong resonance ($1s-2p$) transitions of highly ionized iron, either Fe\,\textsc{xxv} or Fe\,\textsc{xxvi}, but blue-shifted, with an implied outflow velocity of $0.25-0.30$\,c. A column density of $N_{\rm H} \sim 5 \times 10^{23}$\,cm$^{-2}$ was also required to model the high equivalent width ($\sim 100$\,eV) of the lines (R09). Even a conservative 10\% global covering for the ionized outflow implied a mass outflow rate of $>10\,{\rm M}_{\odot}$~yr$^{-1}$, similar to the expected mass accretion rate of PDS 456 \citep{Reeves03}. PDS 456 also exhibits blue-shifted absorption in the UV \citep{O'Brien05}, as seen in the form of a broad absorption trough blue-wards (by $12000-15000$\,km\,s$^{-1}$) of the Lyman-$\alpha$ emission as well as blue-shifted C\,\textsc{iv} emission (by $5000-6000$\,km\,s$^{-1}$).  

PDS\,456 also has a record of showing long-term spectral variability over the last decade of observations (R09, \citet{Behar10}). Indeed in an analysis of all the archived observations of PDS 456, \citet{Behar10} showed the complex variability was likely due to both absorption changes and variations in the intrinsic continuum level, while a relatively invariant ionized reflection component may also be superimposed on the spectra. In some of the observations, the quasar appeared to be highly absorbed, such as in the 2001 \xmm\ observations \citep{Reeves03}, while in the 2007 \xmm\ observations the quasar appeared to be relatively continuum dominated with little intrinsic X-ray absorption \citep{Behar10}.  The overall picture of spectral variability is likely complex, with intrinsic quasar X-ray variability occurring along with the changes in a possible partial covering absorber.

Here we present a comparison between a 2007 and a most recent 2011 \suzaku\ observation of PDS\,456. The original 2007 \suzaku\ observation revealed a steep unabsorbed continuum with photon index $\Gamma>2$, as well as the presence of highly blue-shifted absorption lines at 9\,keV in the quasar rest frame. In contrast the 2011 observation caught the quasar at a substantially lower flux and with a subsequently harder X-ray spectrum. In this paper we present the comparison between the 2007 observation and the 2011 low--flux spectrum and subsequently show that the high velocity outflow is persistent in both the 2007 and 2011 datasets at self-consistent velocities. Furthermore in the 2011 observation, the X-ray spectrum is more likely dominated by reprocessed radiation, possibly from an accretion disk wind. 

Values of H$_{\rm 0}$=70\,km\,s$^{-1}$\,Mpc$^{-1}$, and $\Omega_{\Lambda_{\rm 0}}=0.73$ are assumed throughout and errors are quoted at 90\% confidence ($\Delta\chi^{2}=2.7$), for 1 parameter of interest.

\section{Suzaku Observations of PDS 456}
PDS 456 was first observed by \suzaku\ \citep{Mitsuda07} between 24 Feb to 1 Mar 2007 over a total duration of 370\,ks and with a net (XIS) exposure of 190\,ks after screening. PDS\,456 was subsequently re-observed by \suzaku\ between 16--19 March 2011 for a duration of $\sim240$\,ks, with a corresponding next (XIS) exposure of 125.6\,ks. Both observations were taken with PDS\,456 at the aim-point of the XIS CCD cameras, otherwise known as the XIS nominal mode. A summary of observations is shown in Table\,1. Data were analyzed from the X-ray Imaging Spectrometer XIS \citep{Koyama07} and the PIN diodes of the Hard X-ray Detector HXD/PIN \citep{Takahashi07} and processed usingv2 of the \suzaku\ pipeline. Data were excluded within 436 seconds of passage through the South Atlantic Anomaly (SAA) and within Earth elevation angles or Bright Earth angles of $<5^\circ$ and $<20^\circ$ respectively.

\subsection{XIS analysis}
 XIS data for both observations were selected in $3 \times 3$ and $5 \times 5$  editmodes using grades 0,2,3,4,6, while hot and flickering pixels were removed  using the {\sc sisclean} script. Spectra were (re)-extracted from within  circular regions of 1.5\arcmin\ radius, while the background  was taken from an annulus of $3.6-6.5$\arcmin\ radius centered on the  AGN, but which was free of any background sources. We also checked  an alternate background subtraction method, whereby  background spectra were extracted from 4 circles offset  from the source and avoiding the chip corners containing the calibration sources. Both methods gave identical results,  however we adopted the annulus method as that provided a greater  background area and a more even sampling of the background across the CCDs.  The subsequent background spectra were scaled down  by a factor of $\times 0.075$ to match the size of the extraction region  for the AGN.

 The response matrices ({\sc rmf}) and ancillary response  ({\sc arfs}) files were created using the  tasks {\sc xisrmfgen} and {\sc xissimarfgen}, respectively,   the former accounting for the CCD charge injection and the latter  for the hydrocarbon contamination on the optical blocking filter.  Spectra from the two front illuminated XIS\,0 and XIS\,3 chips  in both observations were  combined to create a single source  spectrum (hereafter XIS--FI) for each observation,  while data from the back illuminated XIS\,1  chip were analyzed separately. Data were included from  0.6--10\,keV for the XIS--FI and 0.6--7\,keV for the XIS\,1 chip. As  the latter is mainly optimized for the soft X-ray band and has  higher background yet small effective area in the iron K band,  we concentrate our analysis here on the XIS--FI spectra.  However we note that the XIS\,0, XIS\,1 and XIS\,3 spectra are  consistent with each other, for each observation (e.g. see R09  for a comparison for the 2007 observation). 

 The net background subtracted source count rates per XIS--FI CCD were  $0.268\pm0.001$\,cts\,s$^{-1}$ (2007) and $0.139\pm0.001$\,cts\,s$^{-1}$  (2011), while the XIS background rates  correspond to only 2.4\% and 3.7\% of the  net source counts for the 2007 and 2011 observations respectively.  The net exposure per XIS--FI CCD for each observation were  190.6\,ks and 125.6\,ks respectively. The XIS spectra and response files were subsequently binned to a minimum energy width corresponding to approximately the HWHM  XIS resolution  of 60\,eV at 6\,keV, dropping to $\sim$ 30\,eV at lower energies.  Channels were additionally grouped to a minimum of 40  counts per energy bin and $\chi^{2}$ minimization  was used for all subsequent spectral fitting. 

\subsection{HXD analysis}
 The extraction of spectra from the non-imaging HXD/PIN instrument  \citep{Takahashi07} is described in detail in R09  for the 2007 \suzaku\ observation of PDS 456.  The same method here is applied to  the 2011 observation. Spectra were re-extracted for both observations over  the 15-50\,keV range,  while the background spectra were extracted from background model D  (as described previously in R09), together with the predicted  Cosmic X-ray Background (CXB) component, using the CXB spectrum and intensity  as measured by \citet{Gruber99}.  The net exposures obtained for  HXD/PIN are summarized in Table\,1. The net source rates obtained  for PDS\,456 were low, of $(1.6\pm0.3)\times10^{-2}$\,counts\,s$^{-1}$ and $<4.5\times10^{-3}$\,counts\,s$^{-1}$ for 2007 and 2011 respectively.  These net count rates represent 3.3\% and $<1.1$\% of the  total background count rate, while the systematic uncertainty of the  background--D model is believed to be typically $\pm1.3$\% at the $1\sigma$ level for a net 20\,ks exposure \citep{Fukazawa09}. Thus PDS \,456 is not detected by HXD/PIN in the subsequent low flux 2011  observation and is marginally detected during the 2007 observation, as  discussed in detail by R09.

 In the subsequent spectral analysis of PDS\,456 presented in Section 3 and 4  we concentrate mainly on the properties of the XIS spectra below 10\,keV,  however we do perform consistency checks of the different spectral models  by comparing the predicted model fluxes in the hard X-ray $15-50$\,keV  band with the measurement or upper-limit deduced during the 2007 and  2011 HXD datasets.

\section{Spectral Analysis} 
Figure 1 shows the overall broad-band  fluxed spectra from 0.6-50\,keV of both observations. Note  the spectra have been unfolded through the instrumental response  against a simple $\Gamma=2$ power-law  model and are not corrected for Galactic absorption.  Below 10\,keV the 2011 observation is clearly harder and shows  continuum curvature down towards the lowest energies, which as discussed below  may be the signature of substantial absorption. In contrast the 2007  spectrum is largely power-law like below 10 keV and appears softer  compared to a photon index of $\Gamma=2$ in Figure 1.

\begin{figure}
\begin{center}
\rotatebox{-90}{
\epsscale{0.75}
\plotone{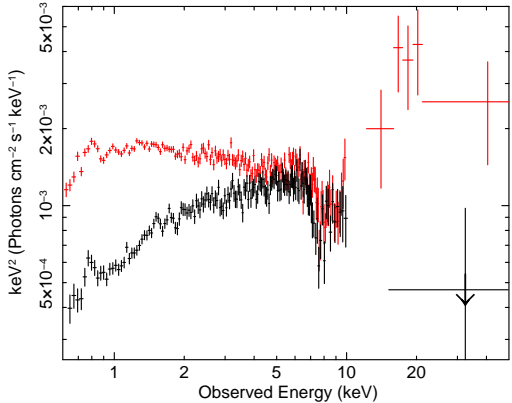}}
\caption{Fluxed spectra of PDS 456, unfolded against a simple $\Gamma=2$ 
power-law.  The Suzaku XIS\,03 and HXD/PIN spectra from 2007 (red) and 2011 (black) are plotted 
against observed energy. Below 10 keV, the XIS spectrum from 2011 is substantially 
harder compared to 2007. Above 15\,keV, PDS 456 is not detected in the HXD/PIN band during 
2011, compared to 2007 where a strong hard X-ray excess is observed. Also note the 
drop in flux in the iron K band above 7\,keV in both observations.}
\end{center}
\end{figure}

Above 10 keV, the 2007  observation shows a clear hard excess in the HXD data, as was originally  presented in R09, while the 2011 observation shows no  such excess. However including the systematic uncertainty in the  HXD background level, then the relative 15-50\,keV fluxes (computed from a  $\Gamma=2$ power-law in the HXD band) are  $(5.7\pm2.2)\times10^{-12}$\,erg\,cm$^{-2}$\,s$^{-1}$ (2007) vs.  $<2.5\times10^{-12}$\,erg\,cm$^{-2}$\,s$^{-1}$ (2011), which are  formally not consistent with being constant at the 90\% level. However  given the possible uncertainties in modeling the non X-ray background  level of the HXD, we do not consider the hard X-ray variability further  in this paper. More sensitive imaging hard X-ray observations  above 10\,keV, such as with NuSTAR, are required  to probe this variability further.  Nonetheless we also note that such hard excesses above 10 keV are proving  to be relatively common, occurring in a high fraction of even type I  AGN \citep{Tatum13}.

For the remainder of the spectral analysis we concentrate on the XIS--FI  data below 10\,keV, which has a very low background level. To simply parameterise the spectral differences below 10\,keV, we started by fitting  both observations simultaneously with a broken power-law model.  Galactic absorption of $N_{\rm H}=2\times10^{21}$\,cm$^{-2}$ \citep{DL90, Kalberla05} was adopted, modeled with the  ``Tuebingen--Boulder'' absorption model ({\sc tbabs} in {\sc xspec}) using  the cross--sections and abundances of \citet{Wilms00}.  The 2007 spectrum is very steep, with soft and hard band photon indices  of $\Gamma=3.16\pm0.07$ and $\Gamma=2.33\pm0.02$ with a break energy of  $1.09\pm0.05$\,keV. In contrast in 2011, the soft and hard band  photon indices are $\Gamma=2.48\pm0.12$ and $\Gamma=1.85\pm0.02$, with  the break energy tied to the above value. The relative soft and hard band  fluxes derived from this model are also quoted in Table 1.

\begin{figure}
\begin{center}
\rotatebox{-90}{
\epsscale{0.75}
\plotone{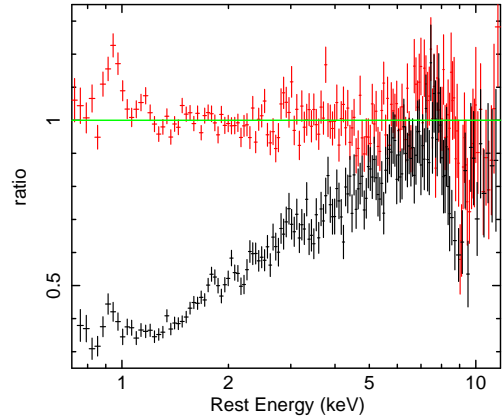}}
\caption{The 2011 Suzaku XIS spectrum of PDS\,456 (black) compared to the 
2007 XIS spectrum. Data are plotted as a ratio to the broken power-law 
continuum model fitted to the 2007 data, as described in the text. 
The spectra are plotted in the rest frame of PDS 456 (at z=0.184). 
Note the strong Fe K band absorption above 8 keV in the rest frame, 
which is especially apparent in the 2011 observation.}
\end{center}
\end{figure}

Nonetheless the fit to the data with such a simple model is very poor,  with $\chi_{\nu}^{2} = 680/314$. The model also does not fit the  pronounced spectral curvature present in 2011, as seen in Figure 1.  Indeed Figure 2 shows the comparison between the 2 datasets, plotted as a  ratio to the absorbed broken power-law with the spectral parameters equal to  the 2007 dataset. In addition to the harder shape  of the 2011 observation, a pronounced spectral drop is observed above 8\,keV  in the quasar rest frame, in the iron K band. Furthermore possible excess  emission is also observed around 0.9-1.0 keV, close to the  expected energies for Ne\,IX or L-shell Fe, as detected previously in this  AGN (R09).   

\subsection{Iron K Band Absorption}
Initially we fitted both observations with a simple power-law continuum model over the 2--10\,keV band,modified by Galactic absorption and allowed the relative photon indices and continuum normalizations to vary for each dataset. Figure 3 shows the data/model ratio residuals at iron K for both observations, plotted in the quasar rest--frame. Indeed the absorption appears to be present in both observations at a self consistent energy between the observations, ruling out the possibility that the iron K band absorption is due to statistical noise. Similar to what was reported by R09 in the analysis of the 2007 dataset, the absorption could consist of two profiles, centered at 9.0 and 9.5 keV respectively in the quasar rest-frame, or alternatively a single broad profile.One apparent difference is that the absorption profile in the 2011 observation appears to extend to lower energies between 8--9 keV. 

\begin{figure}
\begin{center}
\rotatebox{-90}{
\epsscale{0.75}
\plotone{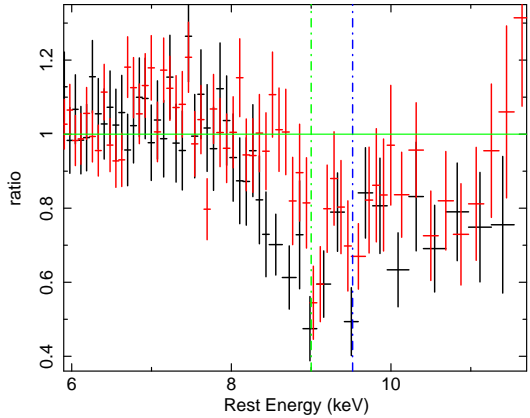}}
\caption{Data/model ratio in the Fe K band for the 2011 (black) and 
2007 (red) observations. Note that the absorption profile is centered 
at a similar rest frame energy in both observations, however the 2011 
absorption profile appears broader and extended red-wards, below 9\,keV. 
The dot--dashed vertical lines are shown as a guide to 
represent the centroid energies of the 
absorption in the 2007 observation.} 
\end{center}
\end{figure}

To quantify its properties, we parameterized the iron K band absorption  with simple Gaussian profiles, fitting two Gaussians lines to the apparent  absorption lines at 9 and 9.5 keV. In addition it was also found that  an additional broad absorption trough was required at a rest frame  energy above 10 keV, which may for instance represent an unresolved blend of  high order transitions or bound-free edges of iron.  The properties of the absorption lines are summarized in Table\,2.  Adding the absorption profiles considerably improved the fit statistic  compared to the baseline power-law model, which decreased from  $\chi_{\nu}^{2} = 410/226$ (formally rejected with a null probability of  $9\times10^{-13}$), to an acceptable $\chi_{\nu}^{2} = 211/215$ upon the addition of the  absorption lines. Thus the detection of the Fe K band absorption  from a statistical sense appears robust, appearing in the same band  in both independent datasets.  In addition R09 also showed that the detection of the  absorption in the 2007 observations was significant at $>99.99\%$  from Monte Carlo simulations, while in a systematic analysis of a  sample of iron K band absorbers in AGN,  \citet{Gofford13} also found that the iron K absorption  profiles in both the 2007 and 2011 observations of PDS\,456  are significant at $>99.9$\% confidence according to Monte Carlo simulations.

The main change in the absorption profile between the two observations is  that the 9 keV absorption line appears centered towards lower energies in  2011, at $8.77\pm0.14$\,keV vs. $9.06\pm0.05$\,keV in 2007, as indicated by Figure 3.  Furthermore the line profile appears to be resolved in 2011 (as can  be seen by the extended absorption profile below $\sim 9$\,keV), with a line  width of $\sigma=0.42^{+0.33}_{-0.12}$\,keV, while in 2007 the profile of the 9\,keV  trough is barely resolved compared with the instrumental resolution  ($\sigma=0.12^{+0.10}_{-0.08}$\,keV). The  equivalent width of the 9 keV absorption profile is also stronger in  2011 compared to 2007. In contrast the 9.5 keV absorption is consistent  across both observations, while the higher energy trough above 10\,keV  also appears to be required in both observations (see Table 2 for details).  Thus overall the Fe K-band absorption appears somewhat stronger in the lower  flux 2011 observation, while the absorption profile is more extended  down towards lower energies. We investigate below a possible cause of the absorption profile variations, from fitting \textsc{xstar} photoionization models. 

\subsection{Line Identifications and Photoionization Models}
Firstly we reconsider the possible identification of the absorption  lines at 9 and 9.5\,keV. As also discussed by R09, if the lines are identified  with the $1s-2p$ transition of Fe\,\textsc{xxvi}, then this implies  a blue-shift of $v_{\rm out}=0.25-0.30c$, or higher still if the lines are  less ionized (He-like or lower). However there are few other likely strong  transitions in the 9-9.5\,keV band that could otherwise account for the  absorption, without a strong velocity shift.  Indeed the closest transition from  iron is the higher order $1s-5p$ line of Fe\,\textsc{xxvi} at 8.91\,keV,  which leads to a series of higher order ($1s-np$, where $n>5$)  lines up until the  the Fe\,\textsc{xxvi} edge at 9.277 keV. However the higher order lines  will usually be substantially weaker than the transitions up to lower  energy levels, unless the lines are on the saturated part of the curve of growth,  which can occur at low turbulence velocities.   Thus at zero (or low) velocity shift, the stronger $1s-2p$ (6.97\,keV)  and $1s-3p$ (8.25\,keV) lines of Fe\,\textsc{xxvi} should be apparent in  Figure 3, which is not the case. We demonstrate this below when we fit  the absorption with a self-consistent grid of photoionized absorber  models.

Alternatively there could be some weaker contribution from other higher Z  elements, namely  Co, Ni and Zn, but these are all substantially under-abundant  astrophysically compared to Fe, e.g. Zn and Co are $\times1000$ less  abundant than Fe \citep{GS98} and would be undetectable with  the current instruments. Ni is somewhat more abundant (but still  typically $\times20$  lower than Fe), however only higher order H or He-like lines of Ni are  expected in the 9-9.5\,keV band. Furthermore in a recent study of iron K absorption with Suzaku,  \citep{Gofford13} found that the effect of Ni was negligible  on the Fe K band spectra computed by \textsc{xstar}.  Thus considering all these possible  alternate identifications for the absorption, the $1s-2p$  resonance absorption lines from iron seem the most plausible identification.

The absorption profiles were fitted with a grid of photoionization models generated by the \textsc{xstar} code v2.2 \citep{Kallman04}. The absorption grid was generated with a turbulence velocity of $\sigma=5000$\,km\,s$^{-1}$, a power-law continuum of photon index $\Gamma=2.2$ and was calculated from $0.10-20$\,keV with 10000 energy bins. The abundances used are from the Solar values of \citet{GS98}. 

The absorption grid provides an excellent fit to the Fe K absorption in both  observations of PDS\,456 with column density  $N_{\rm H}=2.0^{+0.6}_{-0.5}\times10^{23}$\,cm$^{-2}$ and an overall  fit statistic over the 2--10\,keV band of $\chi_{\nu}^{2}=224.1/221$;  see Table\,3 for absorber parameters. Two outflowing zones  are required, with outflow velocities  of $v_{\rm out}=0.25\pm0.01c$ and $v_{\rm out}=0.29\pm0.01c$ respectively,  which can account for the two absorption troughs at 9.0\,keV and 9.5\,keV  respectively. Note that the column density and outflow velocities were  tied to the same values between the two  observations, if they are allowed to vary then the values are formally  consistent at the 90\% level. However there is some evidence  for variability of the ionization parameter between the two observations;  in the lower flux 2011 observation  the absorber is less ionized ($\log\xi=3.44\pm0.06$) vs. the 2007 observation ($\log\xi=3.77^{+0.12}_{-0.14}$). If instead  the ionization parameter is held fixed between the observations, then the  fit statistic is subsequently worse, by $\Delta\chi^2=18.5$ for 1 degree  of freedom. Thus it may be the case that the lower ionisation species of  iron (He-like and lower) contribute towards the lower energy profile of the  absorption in 2011. This is shown in Figure 4 (panel a), where the  fit with a high turbulence grid ($\sigma=5000$\,km\,s$^{-1}$)  and large velocity shift is shown and the extended  lower-energy absorption trough can be seen towards the 2011 observation.

\begin{figure}
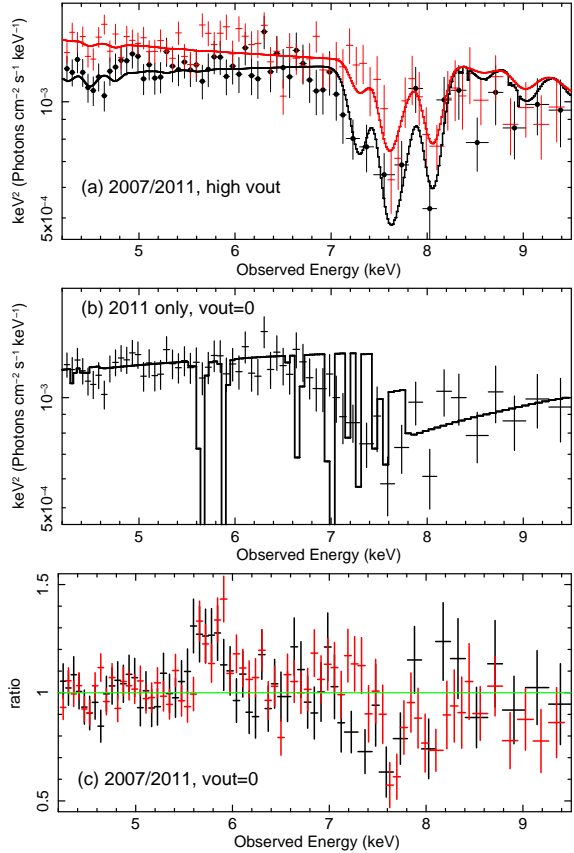

\begin{center}
\rotatebox{-90}{\includegraphics[height=7.5cm]{f4a.eps}}
\rotatebox{-90}{\includegraphics[height=7.5cm]{f4b.eps}}
\rotatebox{-90}{\includegraphics[height=7.5cm]{f4c.eps}}
\end{center}
\caption{\textsc{xstar} fits to the Fe K band absorption profile 
in PDS 456, as described 
in Section 4.1. Panel (a) shows the best fit model with high outflow 
velocity ($v_{\rm out}=0.25-0.29c$) and \textsc{xstar} 
turbulence of $\sigma=5000$\,km\,s$^{-1}$. Panel (b) shows 
the best fit case for zero velocity shift and low turbulence 
($\sigma=100$\,km\,s$^{-1}$). Only the 2011 data are shown 
for clarity. Panel (c) shows the ratio of the low velocity model to 
the 2007 and 2011 data. The 2007 data are shown in red points, 
2011 in black, while models are shown as solid lines. 
Note the plots (a) and (b) were plotted by creating fluxed spectra against a 
simple $\Gamma=2$ power-law and then overlaying the absorption model.} 
\end{figure}

Next, the case where the absorption could be modeled with zero (or small)  velocity shift was investigated. In this scenario an absorption grid  with low turbulence velocity $\sigma=100$\,km\,s$^{-1}$ was used.  In principle this means that for high column densities  some of the usually weaker higher order transitions are of  similar strength to the lower order ones, as the absorption lines lie on the saturated part of the curve of growth (i.e. the opposite of the  high turbulence model). The velocity shift of the absorber was assumed to  be zero. This resulted in a substantially worse fit when  applied to both the 2007 and 2011 spectra, with a fit statistic of  $\chi_{\nu}^{2}=346.3/223$ (vs. $224.1/221$ for the high velocity case),  which is rejected with a null probability of  $2.2\times10^{-7}$. The resulting fit and data/model residuals are  shown in panels (b) and (c) of Figure 4. The column density obtained is  very high, indeed only a lower limit can be obtained of  $N_{\rm H}>2.4\times10^{24}$\,cm$^{-2}$, implying any such gas would be  Compton-thick to electron scattering. Similar to the  high velocity case above, the ionization parameter appears to decrease  from $\log\xi>4.3$ in 2007 to $\log\xi=3.65\pm0.20$ in 2011. 

It is clear that the zero outflow velocity absorption  profile is a poor fit to the data in both the 2007 and 2011 spectra.  In particular the zero velocity model predicts strong $1s-2p$ absorption lines  from He and H-like iron at 6.70 and 6.97 keV respectively (or 5.6--5.9\,keV  observed frame) which are not present in either dataset. This produces an  apparent excess in the data/model ratio residuals in panel (c) of Figure 4,  where the data-points are subsequently under--predicted by the model at those energies.  Furthermore the absorption profile at higher  energies is also inadequately modeled by the blend of narrow, high  order transitions of iron, as is apparent in the residuals. 

Thus both datasets require absorption from highly ionized  species of iron at high outflow velocities, at high statistical confidence.  However there does appear to be subtle changes in the absorption  profile between 2007 and 2011, which may be accounted for  by a decrease in ionization in the lower flux 2011 observation. 

\section{Modeling the Spectral Variability} Having established the robustness of the high velocity outflow in PDS\,456,  we next discuss models to account for the broad-band spectral variability  between the 2007 and 2011 observations. In particular we consider three  scenarios below:- (i) a simple variable intrinsic continuum, (ii) models involving variable partial covering absorption  \citep[e.g.][]{Turner10} or (iii) models  involving Compton reflection \citep{RF93,GK11}.  The spectra are also discussed in  terms of some of the latest radiatively driven disk--wind models, i.e.  using the 2D radiative transfer model of \citet{Sim08}. To model the spectral variability between the two datasets,  we retain the high ionization, high velocity absorber as described  in the previous section (with parameters, $N_{\rm H}$, $v_{\rm out}$ and $\log\xi$  allowed to vary). The continuum  itself is modeled by a power-law of variable normalization and $\Gamma$  however the photon index is assumed not to vary between the observations,  except in the variable  continuum scenario discussed below.  A Gaussian emission line is also required at 0.9\,keV, as  noted previously (R09)  and is included at constant intensity between  observations in all the fits below. The XIS data are modeled over the  energy range from 0.6--10 keV.

\subsection{Intrinsic Continuum Variations}
Before we  consider more complex models for the spectral variability involving  absorption or reflection,   we tested a simple model whereby the primary photon index of the continuum  (along with its normalization) varies. Such a scenario may be related  to changes in the Comptonising electron population responsible for the  hard X-ray power-law \citep[e.g.][]{HM91}.   No other additive (emission) or  multiplicative (absorption) components were included,  other than the  0.9 keV emission line, the high ionization iron K absorber and Galactic absorption.  In this scenario the intrinsic continuum does appear  harder in the 2011 observation, varying from $\Gamma=2.35\pm0.02$ (2007) to  $\Gamma=1.84\pm0.02$ (2011), while the continuum flux varies as per Table 1.  However such a simplistic change results in a statistically poor fit, with  $\chi_{\nu}^{2}=606.4/308$, which is rejected at a very high confidence level  of $1.1\times10^{-21}$. 

\subsection{Partial Covering Models}
Thus in order to successfully account for the variability between  2007 and 2011, we investigated models involving gas that reprocess X-rays,   either from matter that partially obscures and absorbs the AGN emission  or via models that scatter and reflect X-rays off Compton thick matter (see  Section 4.3).  Partial covering models have been successfully invoked many times to explain  complex spectral behaviour  in AGN, e.g. see \citet{TM09} for a review of the phenomenon.  For instance to account for pronounced  continuum curvature below 10\,keV or absorption in the iron K band  \citep[e.g.][]{Tanaka04,Turner05,Reeves05,Miller08}, spectral variability and X-ray occultation  \citep[e.g.][]{Risaliti05,Turner08,Turner11,Behar10,Lobban11,Nardini11,Miyakawa12} or pronounced hard excesses above 10\,keV \citep[e.g.][]{R09,Risaliti09a,Turner10,  Tatum13}. Partial covering  scenarios usually require compact clouds of gas or inhomogeneous structure  within the X-ray absorber, whose size-scales are similar to the typical  extent of the X-ray emission region. Such X-ray absorbing gas may exist in the  form of BLR clouds which may cause at least partial occultation  of the X-ray source \citep[e.g.][]{Lamer03,Risaliti09b,Risaliti09c},  or could represent density fluctuations as part of an outflowing wind  \citep[e.g.][]{PK04,Sim10b}. 

Thus a partial covering model was applied to the 2007 and 2011 observations.  Two layers of partially covering gas are fitted to the spectra, where the model can be expressed phenomenologically as:-

\begin{equation}
{\rm tbabs} \times {\rm xstar}_{\rm Fe} \times {\rm pc}_{2} \times ({\rm po} 
+ {\rm pc}_{1} \times {\rm po})
\end{equation}

\noindent where \textsc{tbabs} represents the ISM absorption  due to our Galaxy, ${\rm xstar}_{\rm Fe}$ is the highly ionized outflowing  absorber responsible for the iron K-shell absorption lines and {\rm po} is  the power-law continuum emission. The components  ${\rm pc}_1$ and ${\rm pc}_2$ represent  the partial covering absorbing layers which have line of  sight covering fractions of $f_1$ and $f_2$ respectively.  The two partial covering zones are required as a  higher column, mildly ionized zone (denoted as ${\rm pc}_1$ above) is needed  to fit the spectral curvature above 2\,keV, as well as a lower column  neutral partial coverer (denoted as ${\rm pc}_2$), in order to model the absorption present in the soft X-ray band which  is most pronounced in 2011. In this scenario, it is assumed that the column density  remains invariant between observations, while the covering fractions,  $f_1$ and $f_2$ are allowed to vary between observations.

The column density of the higher column ${\rm pc}_{1}$ zone is  found to be $N_{\rm H}=(2.1\pm0.3)\times10^{23}$\,cm$^{-2}$,  while the lower column ${\rm pc}_{2}$ gas has  $N_{\rm H}=(1.8\pm0.5)\times10^{22}$\,cm$^{-2}$;  see Table\,4 for full parameter details.  The differences between the two  spectra can then be simply explained by an increase in the covering  fraction of the partial covering absorbers from 2007 to 2011;  for the high column zone,  $f_{1}=0.35^{+0.04}_{-0.05}$ in 2007 increases to $f_{1}=0.54^{+0.11}_{-0.07}$ in 2011, while for the lower column zone,  $f_{2}<0.16$ increases to $f_{2}=0.55\pm0.05$. Interestingly  some net blueshift is required for the high column zone, with  $v_{\rm out}=0.17\pm0.02$c (Table 4),  which may suggest it is connected to the outflowing gas. We  also note that formally this is also consistent with an absorber at $z=0$.  However it would seem less likely that absorption associated  with our own galaxy or local group could partially cover the  line of sight to a distant quasar, while the measured column density  is also $100\times$ higher than the Galactic H\textsc{i} value. 

\begin{figure}
\begin{center}
\rotatebox{-90}{
\epsscale{0.75}
\plotone{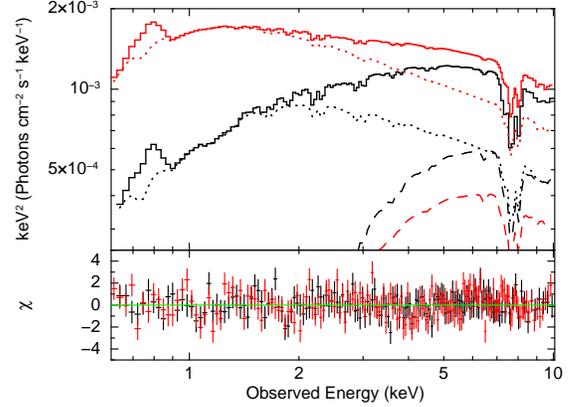}}
\caption{The 2007 (red) and 2011 (black) Suzaku spectra fitted with a partial 
covering model. The upper panel shows the model components, the upper solid lines 
show the total model emission, the dotted lines the unabsorbed power-law continuum and the lower 
dashed lines the absorbed partially covered component. Note the high velocity 
absorber covers both continuum components. The lower panel shows the data/model 
residuals in units of $\sigma$. The partial covering model provides a good fit to the 
data, whereby the relative covering fraction of the absorbed continuum increases from 2007 and 2011.}
\end{center}
\end{figure}

Overall the variable partial covering model provides an excellent fit  to the two observations (see Figure 5), with $\chi_{\nu}^{2}=318/302$. Note that if the $N_{\rm H}$ is also allowed to vary, then it is consistent  between observations.  We also note that correcting for the absorption, the X-ray  luminosity of PDS\,456 differs by very little between the observations, with  $L_{2-10}=4.3-4.6\times10^{44}$\,erg\,s$^{-1}$, which favors the changes  in both spectral shape and continuum flux being caused by absorber  variability in this scenario.

One open question is whether partial covering is really required to explain  the PDS\,456 spectra? One possibility is that the low column, neutral partial  covering zone (${\rm pc}_2$) instead could be described by a partially  ionized (or warm) absorber, similar to those frequently observed in lower luminosity  Seyfert 1 galaxies \citep{Blustin05,McKernan07},  which fully covers the X-ray source. In this case the model  can be expressed as:-

\begin{equation}
{\rm tbabs} \times {\rm xstar}_{\rm Fe} \times {\rm xstar_{\rm WA}} \times ({\rm po} 
+ {\rm pc}_{1} \times {\rm po})
\end{equation}

where ${\rm xstar_{\rm WA}}$ in equation 2  then corresponds to the fully covering warm  absorber zone, of modest ionization, which absorbs the soft X-ray spectrum below 2\,keV.  This description provides an excellent representation of the spectra  (with $\chi_{\nu}^{2}=325/302$). The warm absorber has a  column of $N_{\rm H}=7.0^{+5.0}_{-2.0}\times10^{21}$\,cm$^{-2}$  and ionization parameter  of $\log \xi=1.4\pm0.3$ during the 2011 observation, while no outflow  velocity is required for this absorber. However no such  absorption is required during the softer, less  absorbed 2007 observation, where the column density  of the absorber is constrained to be substantially lower, with  $N_{\rm H}<2.4\times10^{20}$\,cm$^{-2}$. 

Nonetheless the higher column (${\rm pc}_1$) partial coverer is still required  to model the spectra, while its ionization is constrained to be low ($\log\xi<2.1$), with covering fractions as per Table~4.  If this component is removed from the model and the  model refitted, then the fit-statistic is significantly worse with $\chi_{\nu}^{2}=493/306$,  which is rejected with a null hypothesis probability of $5.7\times10^{-11}$.  Thus a variable high column partial coverer would appear to be required  to account for the spectral variability in PDS 456. However one  possibility is that  instead of the AGN being partial covered by absorbing gas, a significant  fraction of the emission is scattered back into our line of sight from  Compton-thick material. 

\subsection{Reflection Models}
Alternatively we investigate whether models involving X-ray reflection  can account for the spectral changes, with the low flux and harder  spectrum in 2011 having a higher fraction of reflected emission, as has been  observed for instance through deconstruction of the time-variable spectra  of Seyfert galaxies through Principal Component Analysis  \citep[e.g.][]{Miller07,Miller08}. We note the lack of a  narrow 6.4\,keV iron K$\alpha$ line rules out a  distant (e.g. pc scale) neutral (or low ionization) reflector,  indeed the upper-limits on the equivalent widths of a narrow 6.4\,keV line are small; $<14$\,eV and $<28$\,eV respectively for 2007 and 2011.  Thus the reflected spectrum is either required to be highly  ionized, or velocity broadened as expected if it originates  from the regions close to the black hole. 

Thus we replace the partial covering (pc1) component of absorption with a table  of X-ray reflection spectra, using the ionized reflection models  recently computed at high resolution by \citep{GK11}, otherwise  known as \textsc{xillver}. We consider the scenario whereby the reflector is highly ionized,  which then results in a large degree of Compton broadening of the iron line.  This then removes the requirement for additional velocity broadening of the  reflection spectrum. However in order to not to over-predict the  emitted soft X-ray flux, the reflection component itself is required to  be absorbed by a layer of photoionized gas. Such a scenario may, to first  order, replicate the reflected and absorbed emission off a disk  wind \citep[e.g.][]{Sim08}, a possibility that we discuss further in Section 4.4.  Thus the model can be expressed as:-

\begin{equation}
{\rm tbabs} \times {\rm xstar}_{\rm Fe} \times {\rm xstar_{\rm WA}} \times ({\rm po} 
+ {\rm xstar}_{\rm ref} \times {\rm reflect})
\end{equation}

where the components are as per equation 2, except the {\rm reflect}  component represents  the reflected emission, which is itself absorbed by a column of  photoionized of gas (${\rm xstar}_{\rm ref}$), with a best fit column of  $N_{\rm H}=2.3\pm0.5 \times 10^{23}$\,cm$^{-2}$ and ionization  parameter $\log \xi < 1.7$. This also provides an acceptable  representation of the 2007 and 2011 spectra of PDS\,456  (with $\chi_{\nu}^{2}=328/302$). The overall model form and relative contributions  of the power-law and absorbed reflection components is very similar phenomenologically  to the partial covering model described in Section 4.2.

The ionization parameter of the reflector is high, $\log \xi>4.1$,  with the dominant  emission occurring from H-like iron. Interesting, there appears to be  a requirement for the reflecting material to be somewhat blue-shifted (or  outflowing), which although is poorly determined, is constrained to be at  least $v_{\rm out}>0.11$\,c. This may also  suggest that the reflecting material  could be associated with an outflowing wind and we note that such an  outflowing reflected component also  appeared to be required in an analysis of the  2007 \xmm\ datasets of PDS\,456 by \citet{Behar10}. For the case  where the bulk velocity of the reflector is forced to be zero, then the  fit statistic is significantly worse by $\Delta\chi^2=33$.  We also note that the strength of the reflected emission is relatively  higher in the low flux 2011 observation compared to the 2007 observation.  Comparing the ratio ($R$) of the  flux of the reflected component to that of the intrinsic power-law  continuum in the 2-10\,keV band gives a ratio of $R=0.60\pm0.16$  during 2011, while in 2007 the ratio is lower with $R=0.32\pm0.09$.  Thus, perhaps as might be expected, the lower flux 2011 spectrum has a relatively  higher fraction of reflected flux, relative to its overall lower intrinsic continuum  level.

\subsection{Disk--wind Models}
In the above sections we showed that relatively ad-hoc absorption and reflection models can  account for the X-ray spectral variability of PDS 456.  Now we consider whether the data  can be fitted self consistently via both absorption through and from reflection off,  the surface of an accretion disk wind.  We adopt a table of synthetic spectra computed for parameterized models of smooth, steady-state 2D bi-conical wind models using the radiative transfer code described by  \citet{Sim08,Sim10a}. The class of wind model used is described in detail by \citet{Sim08,Sim10a} and was also recently  applied to the Suzaku spectra of 6 bare Seyfert 1 galaxies by \citet{Tatum12}.  The computed spectra self consistently contain a combination of the radiation transmitted through the wind and  reflected or scattered emission from the wind (including the iron K$\alpha$  emission).

As discussed in  detail by both \citet{Sim08,Sim10a} and \citet{Tatum12}, the resulting  spectra are strongly dependent on the observer's orientation, which is specified by $\mu = \cos\theta$.  Here, $\theta$ is the angle between the observer's line-of-sight and the polar $z$ axis of the wind, with the  disk lying in the $xy$ plane, e.g. see Figure 6 for a schematic of the  inner wind geometry.

\begin{figure}
\begin{center}
\rotatebox{0}{\includegraphics[height=7.5cm]{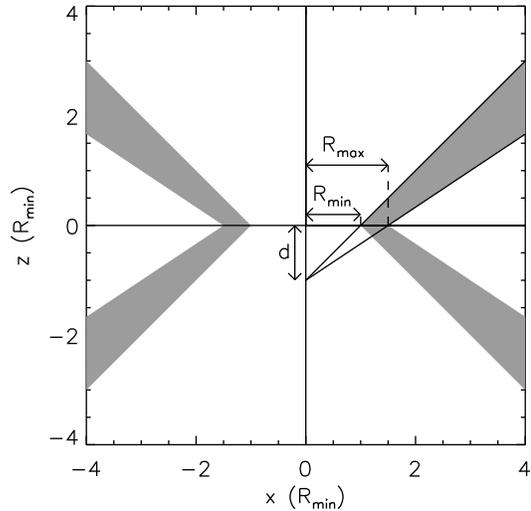}}
\end{center}
\caption{The disk wind geometry employed in Section 4.4, based on the model of 
\citet{Sim08}. 
The x-axis represents the plane of the disk and the z-axis 
the polar direction, in units of $R_{\rm min}$, the minimum launch radius of the flow. 
The black hole is at the origin and inclination angle is measured with 
respect to the z-axis. The shaded area represents the physical extent of the outflow.
The models presented here have $R_{\rm min}=32R_{\rm g}$ along the disk plane, while  
the maximum launch radius $R_{\rm max} = 1.5R_{\rm min}$, as indicated in the figure. 
The d parameter is the distance of the focus point of the wind below the origin in units 
of $R_{\rm min}$ (shown for $d=1$), as indicated by the solid lines. 
Thus increasing the d parameter makes the wind more polar.
Note that only the inner part of the flow is shown for clarity, the flow is computed 
out to a radius of $3.4\times10^{4}R_{\rm g}$.} 
\end{figure}

At low values of $\theta$ (i.e. looking  down at the disk plane; large values of $\mu$)  the resulting spectra have little absorption because radiation from the primary X-ray source, which is assumed assumed to be centrally concentrated, can reach the observer directly without passing through the bi-conical wind structure. However, the spectra do  contain a contribution from reflection off the wind surface which can produce a broadened iron  K line profile. At higher values of $\theta$  (smaller values of $\mu$), the observer sight line passes through the outflow and strong iron K absorption features can be imprinted on the spectrum. Thus as $\mu$ is reduced,  the primary continuum emission becomes increasingly  suppressed by absorption and scattering processes.

Here we adopt a grid of wind model spectra,  computed for a range of wind parameters similar to those described  in \cite{Tatum12}. The grid explores several variable parameters as described below. Primary wind and X-ray source parameters used for our grid of wind models are given in Table~\ref{tab:syn} (all model other model parameters are exactly as described in \cite{Tatum12}.  It is important to note that, for reasons of computational cost, our grid does  not represent a systematic exploration of all of the possible  parameter space of the wind model;  it only varies a sub-set of key model parameters that act as convenient means  by which to vary the typical densities, velocities and ionization conditions  in the flow. This grid is therefore useful as a starting point to investigate whether spectral features can be readily explained by a disk wind. 
 
The grid explores a range of values for both the photon index ($\Gamma$) and luminosity ($L_{\rm X}$) of the illuminating  power-law. The luminosity is calculated in the 2-10\,keV band and  is normalized as a fraction of the Eddington luminosity and thus scales  independently of black hole mass. Models were computed with $L_{\rm X}$ in the range $0.1-1.0$\% of the Eddington luminosity. When fitting the models, the value of  $\Gamma$  was tied to that of the measured photon index of the spectrum, which was  allowed to vary.

Spectra were calculated for two different values of the mass outflow rate of the wind ($\dot{M}$). Normalized to the Eddington accretion rate ($\dot{M}_{\rm Edd}$), the values chosen were   $\dot{M}=\dot{M}_{\rm w}/\dot{M}_{\rm Edd} = 0.32 $ and 0.8, i.e.  where $\log_{10}(\dot{M}_{\rm w}/\dot{M}_{\rm Edd}) = -0.5, -0.1$ respectively, see Table\,5.   Values of the mass outflow rate near to Eddington values  may be expected for PDS\,456, which likely accretes at around the Eddington limit, as discussed in R09. 

The terminal outflow velocities ($v_{\infty}$) realised in the wind models are determined by the choice of the inner wind radius ($R_{\rm min}$) and the terminal velocity parameter  $f_{\rm v}$ which is related the terminal velocity on a wind streamline to the escape velocity at its base, via $v_{\infty} = f_{\rm v} \sqrt{2GM_{\rm BH}/R}$ \citep[see][for details]{Sim08}.  In order for the radial outflow velocities of the wind to reach high values appropriate  for PDS\,456, the spectral models were calculated using a fixed inner radius  of the flow launching region of    $R_{\rm min} = 10^{-0.5} \times 100R_{\rm g} = 32R_{\rm g}$, the same  as the smallest value for the inner radius considered in the \citet{Tatum12}  grid of models.   A relatively low value of the outer radius for the wind launching region of $R_{\rm max} = 1.5 R_{\rm min} = 48 R_{\rm g}$ was also adopted  and thus $R_{\rm min}$ to  $R_{\rm max}$ set the range of radii from which the wind is launched.    We considered models with $f_{\rm v}=1, 2$ (i.e. terminal velocities of one or two times the escape speed from the base of the flow). The opening angles are controlled by the geometrical parameter $d$, which  sets the degree of collimation of the wind: $d$ is   defined as the distance of the focus point of the wind below the origin.  The wind geometry for the innermost part of the flow, along with the  definitions of the $R_{\rm min}$, $R_{\rm max}$ and $d$ parameters, are  illustrated in Figure 6.

For our model grid we considered three values for $d = 1, 2, 5$, expressed in units of $R_{\rm min}$. Here $d=1$ corresponds  to our most equatorial wind (least collimated case) and  $d=5$ correspond to a more polar  wind (most collimated case).   Solar abundances from \citep{Asplund05} were used for all models. For each of our wind models, spectra were computed for 20 different values of the observer inclination parameter $\mu$, uniformly spaced  between $\mu=0.025$ and $\mu=0.975$. In total our grid contains 2160 synthetic spectra.

We first apply the above models to the 2011 Suzaku spectrum of PDS\,456,  as this is the most absorbed spectrum of the two Suzaku observations,  with a strong iron K absorption profile.  Rather than let all the parameters  of the wind vary simultaneously, which could lead to degeneracies in the modeling,  the effect of changing different parameters  on the wind model was systematically explored.  Given the limited number of grid points for both parameters, values of $\dot{M}$ and $f_{\rm v}$ were fixed in each fit. However,   we explored the effect of adopting the different tabulated values by changing the respective parameters in the fits and comparing the  resultant $\chi^2$. Similarly the geometrical $d$ parameter was also kept fixed  at either $d=1, 2$ or 5, as discussed below.  The photon index of the illuminating  power-law was allowed to vary and was assumed to be equal to the power-law  index of the measured spectrum.  The inclination parameter, $\mu$, was also allowed to vary, as  was the 2-10\,keV luminosity $L_{\rm X}$, expressed as a percentage  of $L_{\rm Edd}$. Parameters are measured relative to the redshift  of the quasar at $z=0.184$ throughout. 

The spectrum of PDS\,456 was modeled with a simple power-law attenuated by   Galactic absorption, modified by the wind spectrum which is included in  \textsc{xspec} as a multiplicative table.  The spectrum was also modified by a soft  X-ray warm absorber, as discussed in Section 4.2, which may originate  from gas further away than the putative accretion disk wind.  Alternatively a partial covering absorber also gave an equally good fit to the  soft X-ray absorption. Thus the  overall model is of the form:-

\begin{equation}
{\rm tbabs} \times {\rm xstar_{\rm WA}} \times ({\rm diskwind} \times {\rm po}) 
\end{equation}

Initially the spectrum was modelled with an escape speed parameter of $f_{\rm v} = 1$,  with $d=1$, while the outflow rate was fixed at the lowest value of $\dot{M}=0.32$ (in  Eddington units). The models were fitted over the 0.5--10\,keV range of the XIS 0 and 3  combined spectrum. The resultant fit is plotted in  Figure\,7 (upper panel), with the  resultant best fit wind parameters and reduced $\chi^{2}$   being listed in Table\,6 as Model\,A.   It can be seen that in this case the fit to the iron line profile  is quite poor, as the model fails to reproduce the entire depth or width of the  profile, especially on the blue-shifted side of the profile. This suggests that the velocity of the wind is not sufficiently high to match the  observed profile. Note that increasing either $\dot{M}$ or $d$ (making the outflow  more polar) makes the fit even worse, in the latter case making the velocity profile  shallower. 

\begin{figure}
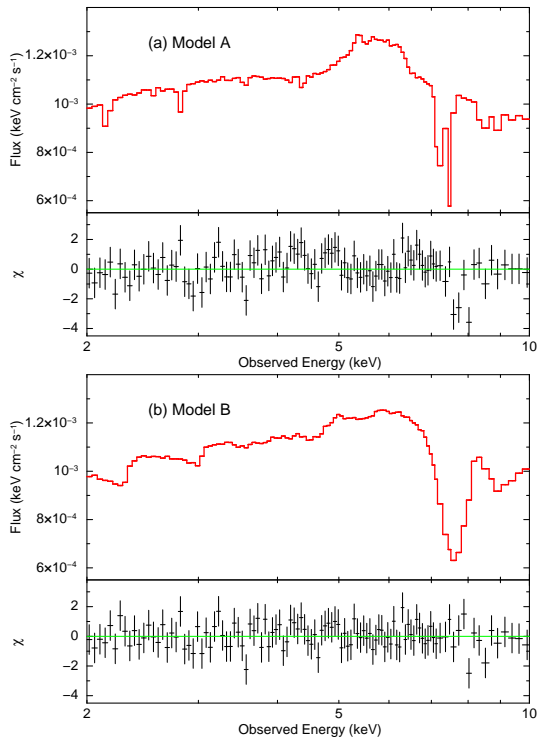

\begin{center}
\rotatebox{-90}{\includegraphics[height=7.5cm]{f7a.eps}}
\rotatebox{-90}{\includegraphics[height=7.5cm]{f7b.eps}}
\end{center}
\caption{The 2011 spectrum of PDS\,456 fitted with the disk-wind model of Sim et al. (2008), the upper panels showing the model and the lower panels showing the residuals to the fit. Model A and B show the fits with two different models, corresponding to two different values of the escape velocity parameter $f_{\rm v}$ equal to 1 or 2 times the escape velocity - see Section 4.4 for details. In each of the fits a low inclination angle ($\mu=\cos \theta = 0.4-0.65$) is required, implying that we may be looking through the marginally Compton-thick part of the wind, see Tables 5 and 6 for model parameters.} 
\end{figure}

Thus, to increase the velocities, the escape velocity parameter was then increased to $f_{\rm v}=2$ in the fits,  while the other parameters are kept  fixed as before, i.e. $\dot{M}=0.32$ and $d=1$. This yields an excellent fit to the  iron absorption profile (Table 6, Model\,B) and is plotted in Figure\,7 (lower panel).  Note that an equally good fit is also obtained if the outflow rate is increased to  $\dot{M}=0.8$, the main difference being that the  percentage 2-10 keV luminosity increases from $\sim0.3$\% to $\sim1$\%. Thus, as might be expected from consideration of ionization conditions, an increase in the ionizing luminosity can be roughly compensated by an increase in the mass outflow rate.

We also investigated the effect of increasing the collimation of the flow, by increasing the  $d$ parameter to $d=2$. This is listed as Model\,C in Table 6, but otherwise the parameters are  the same as for Model\,B and the fit statistic is also similar. In this case the model fit adjusts  via the inclination parameter, which increases from $\mu=0.42\pm0.02$ to $\mu=0.63\pm0.02$.  This also seems consistent with the fact that as the $d$ parameter increases, the outflow becomes  more polar (less equatorial) and thus the inclination angle decreases. However we note that if  the geometric factor is increased further to $d=5$, giving the most polar outflow,  then the fit is no longer acceptable  ($\Delta \chi^2=+21.8$ with respect to Model C), as the absorption feature becomes too shallow.

Thus, in general, a moderately collimated outflow ($d=1-2$), with terminal velocity around twice the escape speed ($f_{\rm v}=2$) for our chosen values of $R_{\rm min}$ and $R_{\rm max}$ appears to model well the absorption profile in PDS\,456. The outflow also appears to be consistent with having a mass-loss rate close to the Eddington accretion rate. It is interesting to note that the value  of the inclination parameter $\mu$ is generally low,  giving $\mu=0.4-0.65$ (or $\theta=55-70\degg$). If the bi-conical wind geometry is correct, this would suggest that  we are viewing PDS\,456 at a relatively side-on inclination through the  thick part of the accretion disk wind.

We caution that our grid does not explore different values for the parameters controlling the extent and location of the wind launching region ($R_{\rm min}$ and $R_{\rm max}$). Thus we cannot yet quantitatively comment on possible degeneracies related to the location of the wind launching region. Since the wind velocities are key to obtaining good agreement, it is to be expected that some degeneracy will exist between the choice of $R_{\rm min}$ and $f_{\rm v}$  (if $R_{\rm min}$ is reduced then a good fit may be achievable for a smaller value of $f_{\rm v}$). While this can be explored by extending our grids of models in the future, we note that it is not expected that either  $R_{\rm min}$ is very much smaller nor $f_{\rm v}$  very much bigger than considered here. 

\subsubsection{Application to the 2007 observation}
For comparison, we also applied the disk wind model described above to the  2007 Suzaku observation. For simplicity, Model B was adopted, with $f_{\rm v}=2$  and $d=1$ and $\dot{M}=0.32$. As the spectrum is brighter and less absorbed, no additional  soft X-ray absorption (warm absorption or partial covering) was applied to the  2007 data, thus the spectrum can be described by an intrinsic power-law  like continuum modified by the presence of the disk wind. The parameters of this fit are also listed in Table 6. The inclination  parameter is consistent with the 2011 model, as might be expected ($\mu=0.43\pm0.02$).  The only change in the model is that the X-ray luminosity increases to the  maximum grid value of 1\% of Eddington ($L_{\rm X}>0.83$, Table 6), while the photon index  is slightly steeper ($\Gamma=2.26\pm0.02$). Thus as the 2007 observation caught PDS 456  in a higher flux than in 2011, it would appear that the ionization state of the wind,  as governed by the luminosity ($L_{X}$) of the illuminating continuum, was also correspondingly  higher. 

\subsubsection{The Physical Conditions within the Wind}
Here we also briefly discuss some of the physical conditions within the  wind model, noting however that the exact parameters are dependent  upon the model fitted and thus should just be viewed as typical estimates.  We also reiterate that there are significant degeneracies among parameters of the wind model fitted here,  which are not adequately explored within the limited parameter space of the current model grid. For reference, however, we take the above best fit case with the wind geometry of Model B, fitted to  both the 2007 and 2011 datasets, for the case of a mass outflow rate (with respect to Eddington) of $\dot{M}=0.32$ and with inclination $\mu=0.42$.  For this model and inclination parameter,   the column density through the flow along the observer's line of sight is estimated to be  $N_{\rm H}\sim2\times10^{24}$\,cm\,$^{-2}$ and thus the Compton depth  is of the order $\tau\sim1$. However the Compton depth through the  flow is strongly inclination dependent; while for low values  of $\mu$ the Compton depth can be reasonably high, reaching $\tau\sim10$  for the case of $\mu=0.05$, for face-on inclinations the depth can  be negligible, e.g. $\tau\sim0$ for $\mu>0.7$ (for the geometry of Model B).  Figure 2 of \citet{Sim08} demonstrates the overall effect of varying inclination on  the Compton depth through the flow.

In the case of Model B, most of the absorption occurs over a physical  radius of typically $\sim 60-160 R_{\rm g}$. This  is likely to be somewhat geometry dependent  and we further note that we fixed the inner and outer radius from which the wind is launched to lie in the range  $32-48R_{g}$. However we can take a reasonable radial scale for the wind  to be of the order $R_{\rm wind}\sim 100R_{g}$, which for PDS\,456,  with an estimated black hole mass of $10^{9}M_{\odot}$ \citep{R09}, would correspond to a radius of  $R_{\rm wind}\sim10^{16}$\,cm.

In that case some typical  physical timescales for the wind can be estimated. A simple light-crossing  or reverberation timescale of the wind would be of the order  $t_{\rm lc}\sim 5\times10^{5}$\,s. The physical flow timescale may be somewhat  (a factor of a few) longer than this, depending on the exact  velocity profile of the wind, although noting that the outflow velocity predicted  from the \textsc{xstar} modeling is fairly fast,  of $v_{\rm out}\sim0.25c$. In contrast the ionization and recombination  timescale of the wind may be fairly short. If, for PDS\,456, we take a  typical column density (depending on the exact line of sight) of  $N_{\rm H}=10^{23}-10^{24}$\,cm$^{-2}$ and a radial length scale  of $10^{16}$\,cm, then the typical electron densities within the flow may  be of the order $10^{7}-10^{8}$\,cm$^{-3}$, although we note that the density in the  wind will not be uniform. This could yield measurable  ionization and recombination times for highly ionized iron within the outflow  (e.g. for Fe\,\textsc{xxvi}) of $t_{\rm rec}\sim10^{4}-10^{5}$\,s \citep{Seaton59}.  Such timescales are certainly well within the 4 years timescale between the  2007 and 2011 observations, thus it would appear reasonable for the ionization  conditions of the wind to have responded to the overall lower flux level in the  2011 observations. Subsequent on-going monitoring of the variability of the wind  in PDS\,456, with \suzaku\ , \xmm\ and {\it NuStar} observations in 2013--2014,  will hopefully provide more insight on the variability timescales.

We can also compare the properties of the fast outflow in PDS\,456, with the  ultra fast outflows typically observed in AGN, such as those  presented in \citet{Tombesi13}.  From the \textsc{xstar} fits shown in Table\,3, the ionization parameter  is typically $\log \xi=3.5-4$ for a column density of  $N_{\rm H}=2\times10^{23}$\,cm$^{-2}$. In that case, PDS\,456 lies in the typical  range of the $N_{\rm H}$ vs. $\log \xi$ distribution for the ultra fast outflows  presented in \cite{Tombesi13}, see their Figure 1, while PDS\,456 lies towards  the higher end of the velocity distribution. This is also the case from comparing  PDS\,456 with the properties of the fast outflows detected in the \suzaku\ sample  of \citet{Gofford13}. For a radial distance scale of the  wind in PDS\,456 of $\log (R/R_{\rm g}) \sim 2$, then according to the radial correlations in  \citet{Tombesi13}, we are viewing the  wind in PDS\,456 along the innermost measured  part of the AGN outflow (see their Figure 3 of ionization,  column or outflow velocity vs radial location in the wind). Thus PDS\,456  would appear to have the typical properties of an ultra fast outflow, although  we could be viewing the quasar through the innermost part of the disk wind.

Finally we can make an approximate estimate for the sizescale of the  soft X-ray absorbing region, which can either be associated  with the more typical warm absorbing regions found in Seyfert 1s  \citep{Blustin05, McKernan07},  or with near neutral matter partially covering the X-ray source. Although  no velocity information on the soft X-ray absorber is available from  these data, if it is associated with the typical velocities found in  soft X-ray warm absorbers, of $\sim 1000$\,km\,s$^{-1}$, then for  a variability timescale within 4 years,  the implied absorber sizescale is $\sim10^{16}$\,cm. Although the  estimate is very much dependent on the absorber velocity, this sizescale  is similar to the fast absorber associated with the  accretion disk wind.

\subsection{Is PDS 456 X-ray quiet?}
One final consideration is how much the intrinsic X-ray emission  in PDS\,456 is suppressed  by the disk wind, especially as it appears that PDS\,456 may be directly  viewed through the marginally Compton thick part of the flow at moderate inclination.  The observed 2-10\,keV luminosity of PDS\,456 in 2011 is  $L_{2-10}=2.6\times10^{44}$\,erg\,s$^{-1}$. This would make PDS\,456  relatively X-ray quiet compared to bright radio-quiet quasars \citep{Elvis94},  as the observed 2-10 X-ray luminosity is only $\sim0.3$\%  of the total bolometric luminosity of  $L_{\rm bol}\sim10^{47}$\,erg\,s$^{-1}$ \citep{Simpson99,O'Brien05,R09}.  Indeed such a low 2-10\,keV luminosity is also predicted   by the disk wind modeling, as the fit typically favoured $L_{\rm X}$ to be typically  about $\sim0.3$\% of the Eddington luminosity,  which for the estimated black hole mass of $\sim 10^{9} M_{\odot}$ (R09),  is $L_{\rm Edd} \sim 10^{47}$\,erg\,s$^{-1}$ for PDS\,456. 

We note that the X-ray flux of PDS\,456 is attenuated by the disk  wind by a factor of $\times 2$ in the models fitted here, which could in part  explain the X-ray weakness. Potentially further attenuation of the X-ray emission could occur if  shielding gas were also present in PDS\,456, which is not accounted for in the  \citet{Sim08} model, but may be needed for the wind not to become  over ionized in the acceleration region \citep{Murray95}.  The partial covering absorber discussed in previous sections  could possibly act as a source of such material if it is located near the base of the  wind, which would in effect lower the X-ray luminosity that both the wind and  the observer sees. PDS\,456 may also be similar in this regard to BAL quasars, which are known to be X-ray weak due to absorption  \citep{Gallagher02}.

Alternatively it may be that PDS\,456 is intrinsically X-ray weak, as the observed  bolometric correction for the 2-10\,keV band luminosity is a factor $\times 300$,  much higher than in typical AGN \citep{Elvis94}. It is possible, however, for the 2-10\,keV   bolometric correction to reach values of $\sim 100$ for Eddington limited sources \citep[e.g.][]{VF09} and thus AGN like PDS 456  may therefore be intrinsically X-ray weak. Although PDS\,456  is a high luminosity AGN, it may have  some similarities with Narrow Lined Seyfert 1s, which can also have weak  2--10\,keV emission with respect to bolometric, due to their  steep X-ray photon indices \citep{Leighly99}. While PDS\,456  has broad Balmer lines (e.g. H$\beta$ FWHM $\sim3500$\,km\,s$^{-1}$),   like the NLS1s it does have weak [O\,\textsc{iii}] yet strong  Fe\,\textsc{ii} emission \citep{Simpson99},  while its intrinsic X-ray photon index is also  steep ($\Gamma=2.4$). Thus PDS\,456 may also be a high luminosity  analogue of the NLS1s, with a high Eddington ratio.

\section{Conclusions}
We have presented a new 2011 \suzaku\ observation of the nearby,  luminous quasar PDS 456.  The new observation has confirmed the detection of the  iron K-shell absorption  lines that were detected in the 2007 \suzaku\ observation \citep{R09},  at $\sim9$\,keV in the quasar rest frame.  Indeed the possibility of highly blueshifted absorption in  PDS\,456 was first suggested by \cite{Reeves03} as evidence for a fast, powerful wind.  We have shown that the 9\,keV band absorption cannot be modeled with  an absorber with zero (or small) outflow velocity and instead is best  modeled by highly blue-shifted lines from Fe\,\textsc{xxv-xxvi},  with an outflow velocity of $0.25-0.29c$. This makes PDS\,456 one of the  fastest high velocity outflows known, compared to those detected in recent X-ray  samples \citep{Tombesi10,Gofford13}, which tend to cover the range from $0.01-0.30c$.

The overall X-ray spectrum of PDS 456 in 2011 also appears to be harder and observed at an  lower overall flux, compared  to the 2007 observation. This spectral variability could not simply be accounted for  by a simple change in continuum level and photon index between the observations.  The differences between the two observations can be explained by variations in  the line of sight covering fraction of a partial covering absorber. This interpretation  has also been favored to explain the X-ray spectral variability of several type I AGN, which  can sometimes display low and hard absorbed spectra, which may be  due to partial occultation by absorbing clouds \citep[e.g.][]{Risaliti05,Turner08,Turner11,Behar10,Lobban11 Nardini11,Miyakawa12,Pounds13}.  Alternatively the 2011 spectrum may also contain a higher  fraction of reprocessed and scattered X-ray emission, for instance off the surface  of an accretion disk wind. 
 Alternatively the 2011 spectrum may have a higher  fraction of reprocessed and scattered X-ray emission, for instance off the surface  of an accretion disk wind.  Indeed we have shown that the strongly blue-shifted Fe K band absorption in PDS\,456 can be produced by a simple smooth 2D parameterization of an  accretion disk wind \citep{Sim08}. In PDS 456, it may appear that we are viewing the  quasar at relatively low inclination through at least a marginally Compton  thick part of an accretion disk wind, which can also explain the absorbed  nature of the X-ray spectrum, especially at a lower flux during 2011.  PDS\,456 also shows characteristics similar to BAL quasars and is also X-ray weak,  which may be accounted for by enhanced X-ray absorption towards the quasar.  The iron K band outflow velocity of PDS\,456 is also similar  to the X-ray outflow velocity measured in the well studied BAL quasar, APM 08279+5255,  at $z=3.97$ \citep{Chartas02}.  Interestingly an earlier UV snapshot spectrum of PDS\,456 \citep{O'Brien05}, with HST/STIS,  revealed the  possibility of broad absorption troughs, e.g. at the Lyman-$\alpha$ line, as well as  a highly blueshifted C\,\textsc{iv} emission profile. Further upcoming spectroscopy with  HST/COS will hopefully shed light on the possible UV BAL like features in PDS\,456.

  Generally, accretion disk driven winds appear to be a promising  starting point to explain the increasing  number of ultra fast outflows now been observed in AGN X-ray spectra  \citep{Tombesi10,Tombesi11,Gofford13}. Such profiles, such as the smooth 2D  wind model of \citet{Sim08}, can also self consistently explain their  iron line emission profiles as well as absorption and can reproduce an  apparent P-Cygni like profile, as measured for instance in PG\,1211+143 \citep{Pounds09,Sim10a}. Disk winds models can also plausibly reproduce  the broad iron lines of some Seyfert 1s, through reflection off the wind surface \citep{Tatum12}.  In order to accurately quantify such winds, more realistic  geometries (likely including sub-structure/clumping) need to be explored, as well as  time-variability in the flow; indeed a radiatively driven AGN disk wind might be  expected to  display complex time dependence \citep{Proga00,PK04}.  Further deep observations of PDS\,456, scheduled  with \suzaku\ and \xmm during 2013,  will further explore the short time-scale variability  of the wind spectrum. In the future, iron K absorption profiles  will be revealed with unprecedented resolution with the SXS calorimeter on-board {\it Astro-H}.

\section{Acknowledgements}

J.N. Reeves acknowledges financial support from STFC and {\it Chandra} grant number GO1-12143X. T.J. Turner acknowledges NASA grant number NNX11AJ57G. We would like to thank Javier Garcia, for the use of his \textsc{xillver} reflection model. This research has made use of data obtained from the {\it Suzaku} satellite, a collaborative mission between the space agencies of Japan (JAXA) and the USA (NASA).

\begin{deluxetable}{lccccccc}
\tablecaption{Summary of PDS 456 Suzaku Observations}
\tablewidth{0pt}
\tablehead{
	\colhead{Instrument} & 
	\colhead{Year} & 
	\colhead{Start Date/Time$^{a}$} & 
	\colhead{End Date/Time$^{a}$} & 
	\colhead{Exposure$^{b}$} & 
	\colhead{$F_{0.5-2}$$^{c}$} & 
	\colhead{$F_{2-10}$$^{d}$} & 
	\colhead{$F_{15-50}$$^{e}$}}

\startdata

XIS & 2007 & 2007/02/24 17:58 & 2007/03/01 00:51 & 190.6 & 3.46 & 3.55 & -- \\
HXD/PIN & 2007 &  -- & -- & 164.8 & -- & -- & $5.7\pm2.2$ \\
XIS & 2011 & 2011/03/16 15:00 & 2011/03/19 08:33 & 125.6 & 1.36 & 2.84 & -- \\ 
HXD/PIN & 2011 & -- & -- & 93.6 & -- & -- & $<2.5$ \\

\enddata

\tablenotetext{a}{Observation Start/End times are in UT.} 
\tablenotetext{b}{Net exposure time, after screening and deadtime correction, in ks.}
\tablenotetext{c}{Observed flux in the 0.5-2\,keV band, units $\times10^{-12}$\,erg\,cm$^{-2}$\,s$^{-1}$}
\tablenotetext{d}{Observed flux in the 2-10\,keV band, units $\times10^{-12}$\,erg\,cm$^{-2}$\,s$^{-1}$}
\tablenotetext{e}{Observed flux in the 15-50\,keV band, units $\times10^{-12}$\,erg\,cm$^{-2}$\,s$^{-1}$}

\end{deluxetable}


\begin{deluxetable}{llccc}
\tablecaption{Gaussian Iron K Absorption Line Parameters.}
\tablewidth{0pt}
\tablehead{
\colhead{Component} & \colhead{Parameter} & \colhead{2007 values} & \colhead{2011 values} & 
\colhead{$\Delta \chi^{2}$$^{a}$}}

\startdata

Line\,1 & Energy (keV) & $9.06\pm0.05$ & $8.77\pm0.14$ & 124.3 \\
& $\sigma$\,(keV) & $0.12^{+0.10}_{-0.08}$ & $0.42^{+0.33}_{-0.12}$ \\
& Flux$^{b}$ & $-(3.2\pm0.8)\times10^{-6}$ & $-(9.7\pm2.2)\times10^{-6}$ \\
& EW$^{c}$\,(eV) & $-129\pm32$ & $-370\pm84$ \\
\hline
Line\,2 & Energy (keV) & $9.54\pm0.06$ & $^{t}$ & 37.6 \\
& $\sigma$\,(keV) & $0.1^{f}$ & $0.1^{f}$ \\
& Flux$^{b}$ & $-(2.5\pm0.9)\times10^{-6}$ & $-(3.2\pm1.2)\times10^{-6}$ \\
& EW$^{c}$\,(eV) & $-121\pm44$ & $-156\pm59$ \\
\hline
Line\,3 & Energy (keV) & $10.5\pm0.2$ & $^{t}$ & 36.9 \\
& $\sigma$\,(keV) & $>0.3$ & $^{t}$ \\
& Flux$^{b}$ & $-(3.3\pm1.6)\times10^{-6}$ & $-(5.8\pm2.0)\times10^{-6}$ \\
& EW$^{c}$\,(eV) & $-180\pm87$ & $-296\pm102$ \\
\hline
Power-law & $\Gamma$ & $2.23\pm0.03$ & $1.84\pm0.04$ \\
& Flux$^{d}$ & 3.55 & 2.77 \\
\enddata

\tablenotetext{a}{Improvement in $\chi^{2}$ upon adding component to model.} 
\tablenotetext{b}{Line flux in units of photons\,cm$^{-2}$\,s$^{-1}$.}
\tablenotetext{c}{Absorption line equivalent width.}
\tablenotetext{d}{Continuum flux in the 2-10\,keV band, units $\times10^{-12}$\,erg\,cm$^{-2}$\,s$^{-1}$.}
\tablenotetext{f}{Parameter fixed in model.}
\tablenotetext{t}{Parameter is tied between 2007 and 2011 spectra.}
\end{deluxetable}

\begin{deluxetable}{lcc}
\tablecaption{Xstar Model Parameters to Iron K Absorption}
\tablewidth{0pt}
\tablehead{
\colhead{Parameter} & \colhead{2007 values} & \colhead{2011 values}}

\startdata
$N_{\rm H}$$^{a}$ & $2.0^{+0.6}_{-0.5}$ & $^{t}$ \\
$\log \xi$$^{b}$ & $3.76^{+0.12}_{-0.14}$ & $3.44\pm0.06$ \\
$v_{\rm out1}$$^{c}$ & $0.25\pm0.01c$ & $^{t}$ \\
$v_{\rm out2}$$^c$ & $0.29\pm0.01c$ & $^{t}$
\enddata

\tablenotetext{a}{Hydrogen column density in units of $\times10^{23}$\,cm$^{-2}$.} 
\tablenotetext{b}{Log ionization parameter. Units of $\xi$ are erg\,cm\,s$^{-1}$.}
\tablenotetext{c}{Outflow velocity, units $c$.}
\tablenotetext{t}{Parameter is tied between 2007 and 2011 spectra.}
\end{deluxetable}

\begin{deluxetable}{llcc}
\tablecaption{Parameters of Partial Covering Absorption}
\tablewidth{0pt}
\tablehead{
\colhead{Component} & \colhead{Parameter} & \colhead{2007 values} & \colhead{2011 values}}

\startdata
pc1 & $N_{\rm H}$$^{a}$ & $21.0\pm2.5$ & $^{t}$\\
& $\log \xi$$^{b}$ & $<2.1$ & $^{t}$\\
& $v_{\rm out}$$^{c}$ & $(0.17\pm0.02)c$ & $^{t}$\\
& $N_{\rm abs}$$^{d}$ & $1.3\pm0.2$ & $1.9^{+0.7}_{-0.3}$\\
& $N_{\rm unabs}$$^{e}$ & $2.4\pm0.1$ & $1.6\pm0.2$\\
& $f_{\rm cov}$$^{f}$ & $0.35\pm0.05$ & $0.54^{+0.11}_{-0.07}$\\ 
\hline
pc2 & $N_{\rm H}$$^{a}$ & $1.8\pm0.5$ & $^{t}$\\
& $f_{\rm cov}$$^{f}$ & $<0.16$ & $0.55^{+0.05}_{-0.03}$\\ 
\hline
Power-law & $\Gamma$ & $2.50^{+0.11}_{-0.04}$ & $^{t}$
\enddata

\tablenotetext{a}{Hydrogen column density, units $\times 10^{22}$\,cm$^{-2}$.} 
\tablenotetext{b}{Log ionization parameter. Units of $\xi$ are erg\,cm\,s$^{-1}$.}
\tablenotetext{c}{Outflow velocity, units $c$.}
\tablenotetext{d}{Normalization of absorbed power-law. Units are 
in photons\,cm$^{-2}$\,s$^{-1}$\,keV$^{-1}$ at 1\,keV.}
\tablenotetext{e}{Normalization of unabsorbed power-law. Units are 
in photons\,cm$^{-2}$\,s$^{-1}$\,keV$^{-1}$ at 1\,keV.}
\tablenotetext{f}{Covering fraction of partial coverer.}
\tablenotetext{t}{Parameter is tied between 2007 and 2011 spectra.}
\end{deluxetable}

\begin{deluxetable}{lccc}
\tablecaption{Parameters for Disk Wind Model Grid} 
\tablewidth{0pt}
\tablehead{
\colhead{Parameter} & \colhead{min value} & \colhead{max value} & \colhead{grid points}}
\startdata
$\log_{10} (\dot{M}_{\rm w}/\dot{M}_{\rm Edd})$ & -0.5 & -0.1 & 2\\
$\log_{10} (R_{\rm min}/100 R_{\rm g})$ & -0.5 \\
$\mu=\cos\theta$ & 0.025 & 0.975 & 20 \\
$d/R_{\rm min}$ & 1 & 5 & 3\\
$R_{\rm max}/R_{\rm min}$ & 1.5 \\
$f_{\rm v}$ & 1 & 2 & 2 \\
Fe abundance/Solar & 1.0 \\
$\Gamma$ & 1.8 & 3.0 & 3\\
$L_{\rm X}/L_{\rm Edd}$ & 0.001 & 0.01 & 3 \\
\enddata 
\tablecomments{The parameters varied in the grid of wind models, 
listing minimum value, maximum value and number of grid 
points where applicable. See Section 4.4 for details of model parameters.}
\label{tab:syn}
\end{deluxetable}

\begin{deluxetable}{lcccc}
\tablecaption{Disk Wind Model Parameters Fitted to the Suzaku spectrum.}
\tablewidth{0pt}
\tablehead{
\colhead{Parameter} & \colhead{Model A (2011)} & \colhead{Model B (2011)} & 
\colhead{Model C (2011)} & \colhead{Model B (2007)}} 

\startdata
$f_{\rm v}^{a}$ & 1 & 2 & 2 & 2\\
$d^{b}$ & 1 & 1 & 2 & 1\\
$\dot{M}^{c}$ & 0.32 & 0.32 & 0.32 & 0.32\\
$L_{\rm X}^{d}$ & $0.29\pm0.08$ & $0.34\pm0.07$ & $0.28\pm0.06$ & $>0.83$ \\
$\mu^{e}$ & $0.66\pm0.02$ & $0.42\pm0.02$ & $0.63\pm0.02$ & $0.43\pm0.02$ \\
$\Gamma$ & $2.15\pm0.05$ & $2.05\pm0.06$ & $2.10\pm0.05$ & $2.26\pm0.02$ \\
$\chi^{2}/{\rm dof}$ & 163.1/148 & 128.9/148 & 136.1/148 & 217.4/162 \\ 
\enddata

\tablenotetext{a}{Model values for escape velocity parameter 1 or 2 times 
the escape speed, see Section 4.4.} 
\tablenotetext{b}{Geometrical (collimation) factor $d$, see Section 4.4 for details.}
\tablenotetext{c}{Mass outflow rate in units of the Eddington rate.}
\tablenotetext{d}{2-10\,keV luminosity as a percentage of the Eddington luminosity.}
\tablenotetext{e}{Inclination angle parameter $\mu=\cos \theta$.}
\end{deluxetable}

\end{document}